\documentclass[pra,aps,amsmath,amssymb,amsfonts,twocolumn,nofootinbib,longbibliography,superscriptaddress]{revtex4-2}
\usepackage{amssymb}
\usepackage{bm,mathrsfs}
\usepackage{graphicx}
\usepackage{subfigure}
\usepackage{epsfig}
\usepackage{amsmath,bbm}
\usepackage{amsfonts,amssymb}
\usepackage{times}
\usepackage{verbatim}
\usepackage{diagbox}
\usepackage[sort&compress]{natbib}
\usepackage{amsmath}
\usepackage{multirow}
\usepackage[colorlinks=true,citecolor=blue,linkcolor=blue,urlcolor=blue]{hyperref}
\usepackage[usenames]{color}

\newcommand{\cw}{\circlearrowright}
\newcommand{\ccw}{\circlearrowleft}

\usepackage{amsmath}
\DeclareMathOperator\arctanh{arctanh}
\usepackage{float}
\usepackage{graphicx}
\usepackage{svg} 
\usepackage{overpic}
\UseRawInputEncoding
\usepackage{tabularx}

\begin{document}

\title{Magnonic entanglement in a chiral cavity-magnon coupling system}
\author{Yuxin Kang}
\affiliation{Center for Quantum Sciences and School of Physics, Northeast Normal University, Changchun 130024, China}

\author{Xin Zeng}
\affiliation{Center for Quantum Sciences and School of Physics, Northeast Normal University, Changchun 130024, China}

\author{Wuji Zhang}
\affiliation{Center for Quantum Sciences and School of Physics, Northeast Normal University, Changchun 130024, China}
\author{Chunfang Sun}
\affiliation{Center for Quantum Sciences and School of Physics, Northeast Normal University, Changchun 130024, China}

\author{Chunfeng Wu}
\affiliation{Science, Mathematics and Technology, Singapore University of Technology and Design, 8 Somapah Road, Singapore 487372, Singapore}

\author{Gangcheng Wang}
\email{wanggc887@nenu.edu.cn}
\affiliation{Center for Quantum Sciences and School of Physics, Northeast Normal University, Changchun 130024, China}

\date{\today}
\begin{abstract}
The generation of magnon entanglement and squeezing plays a crucial role in quantum information processing. In this study, we propose a scheme based on a chiral cavity-magnon system, which consists of a torus-shaped cavity and two yttrium iron garnet spheres. The magnon mode of each yttrium iron garnet sphere is selectively coupled to one of the two degenerate rotating microwave modes of the toroidal cavity. The system aims to achieve entangled and squeezed magnon states through the mediation of the cavity. We further show that bipartite entanglement can be achieved by tuning external driving parameters. Additionally, our scheme does not rely on the magnon Kerr nonlinearity, which is usually extremely weak in yttrium iron garnet spheres. This work provides insights and methods for the research of quantum states in cavity-magnon systems.
\end{abstract}

 \maketitle
\section{Introduction}
\label{SecI}

Quantum entanglement is a fundamental concept in quantum physics and has garnered significant interest in quantum technologies due to its potential applications in quantum information processing \cite{RevModPhys.81.865}. In continuous-variable (CV) systems \cite{RevModPhys.84.621}, entanglement is intrinsically linked to squeezing \cite{Walls1983}. Squeezing is a crucial resource for generating entanglement in CV Gaussian systems \cite{PhysRevA.64.063811,PhysRevA.65.032323,PhysRevA.66.024303}. Its importance extends to various applications in quantum optics and CV quantum information, including quantum metrology \cite{Schnabel2010,Abadie2011,Aasi2013}, secure quantum communication \cite{PhysRevA.63.052311,PhysRevA.61.010303,PhysRevA.61.022309,PhysRevA.62.062308}, quantum teleportation \cite{PhysRevLett.80.869,PhysRevA.60.937}, cluster states \cite{PhysRevLett.98.070502,Su:12}, heralded gates \cite{Zhao2020}, and quantum computation \cite{doi:10.1126/science.aay4354,doi:10.1126/science.aay2645}. To achieve these applications, researchers have proposed various methods for generating squeezed states. Several experimental schemes for generating squeezed states have been implemented in quantum platforms such as optical \cite{PhysRevA.81.015804,PhysRevLett.118.083604,PhysRevLett.131.133402} and optomechanical systems \cite{PhysRevA.89.063805,PhysRevA.87.063846,PhysRevLett.110.253601}. 

Notably, yttrium iron garnet (YIG) spheres have emerged as a promising platform for macroscopic quantum phenomena and quantum technologies, benefiting from their high spin density, strong exchange interactions, and ultralow loss rates (1 MHz) \cite{PhysRevLett.111.127003, PhysRevLett.113.083603, PhysRevApplied.2.054002, PhysRevLett.113.156401}. A hybrid quantum system that exploits collective spin excitations in YIG spheres has enabled the observation of strong to ultrastrong magnon-photon coupling, mediated by magnetic dipole interaction between magnetostatic modes in YIG spheres and microwave cavity modes\cite{PhysRevLett.113.083603, PhysRevLett.113.156401, PhysRevLett.128.047701, PhysRevApplied.2.054002, PhysRevLett.120.057202, PhysRevB.94.054433}. This hybrid system has been used to investigate various intriguing phenomena, including the generation of quantum states \cite{PhysRevLett.121.203601,PhysRevResearch.1.023021,PhysRevA.99.021801,10.1063/5.0015195,PhysRevLett.124.053602,PhysRevB.103.224416,PhysRevA.101.042331,PhysRevB.100.134421,PhysRevA.101.063838,PhysRevA.103.052411,PhysRevB.105.094422,PhysRevResearch.3.013192}, nonreciprocity \cite{PhysRevLett.123.127202,PhysRevApplied.12.034001,PhysRevA.106.053714}, non-Hermitian physics \cite{Zhang2017,PhysRevA.103.063708,PhysRevLett.121.137203}, and Floquet engineering \cite{PhysRevLett.125.237201,PhysRevA.106.012609,PhysRevA.106.033711,Xie_2023}, thus opening up new possibilities for the generation of squeezed states. For example, in a hybrid cavity-magnon-qubit system, macroscopic entangled states \cite{PhysRevLett.121.203601} and squeezed states \cite{PhysRevA.99.021801,PhysRevA.108.063703} can be prepared by utilizing the nonlinear magnetostrictive interaction. Another approach involves using external quantum drivers, such as single-mode or two-mode squeezed vacuum fields, to entangle two magnon modes in two macroscopic YIG spheres \cite{10.1063/5.0015195,Yu_2020}. These magnon modes can also be entangled with a mechanical mode \cite{Li_2021}, enabling the control of unidirectional quantum steering \cite{PhysRevApplied.15.024042,ZHONG2021127138,Zhang:22}. However, they inevitably introduce nonlinearity into the system or require the injection of external quantum resources.  On the other hand, as the fundamental excitation of spin waves, the magnon inherits the ability of single-spin precession, consequently exhibiting chirality \cite{PhysRevB.85.020406}. This chiral effect is particularly noticeable in the coupling between magnons and optical fields \cite{PhysRevApplied.13.044039, PhysRevB.102.064416, fan2024nonreciprocalentanglementcavitymagnomechanics, PhysRevApplied.19.014030, PhysRevA.108.033701}. For example, in high-quality cavities (such as the WGM microresonator \cite{PhysRevLett.110.213604, PhysRevLett.111.193601, Sedlmeir:13} or waveguides \cite{PhysRevLett.107.173902}), when a YIG sphere is in the Kittel mode \cite{PhysRev.73.155, Lachance-Quirion_2019}, the magnetization rotates counterclockwise \cite{YU20231} around the effective magnetic field and couples with photons of the same polarization \cite{PhysRevLett.124.107202, PhysRevB.101.094414, PhysRevApplied.13.044039, PhysRevApplied.19.014030, PhysRevB.102.064416}. During this process, the polarization of the light in the cavity mode becomes ``locked'' to a specific plane or propagation direction \cite{Lodahl2017, doi:10.1126/science.aaa9519, PhysRevA.90.043802, PhysRevB.101.094414, PhysRevLett.124.107202, Shao2018}, meaning that the polarization direction is closely related to its linear momentum. Therefore, this hybrid magnon system holds promise as a platform for exploring chiral quantum optical phenomena \cite{PhysRevA.101.043842, PhysRevApplied.16.064066, PhysRevA.105.013711, PhysRevA.103.053501, PhysRevB.106.104432}.

In this work, we proposed a scheme to generate a pair of entangled YIG spheres without relying on any nonlinear effects or externally applied squeezed microwave fields. Our strategy is to utilize chiral cavity-magnon coupling. The proposed method circumvents the complexities associated with nonlinearities and external squeezing fields, offering novel insights into exploring macroscopic quantum effects and advancing quantum information processing in cavity-magnon systems

The structure of this paper is organized as follows. In Sec. \ref{Sec.II}, we present a comprehensive description and analysis of the proposed chiral cavity-magnon model and its Hamiltonian. Sec. \ref{Sec.III} describes the Heisenberg-Langevin equation to characterize the system dynamics, introducing two distinct protocols for generating two-mode squeezed states between two magnons and addresses the experimental feasibility of the proposed chiral cavity-magnon system. Finally, concluding remarks summarizing the research findings are provided.

\section{The model and equation of motion}
\label{Sec.II}

\begin{figure}
\centering
\includegraphics[width=0.45\textwidth]{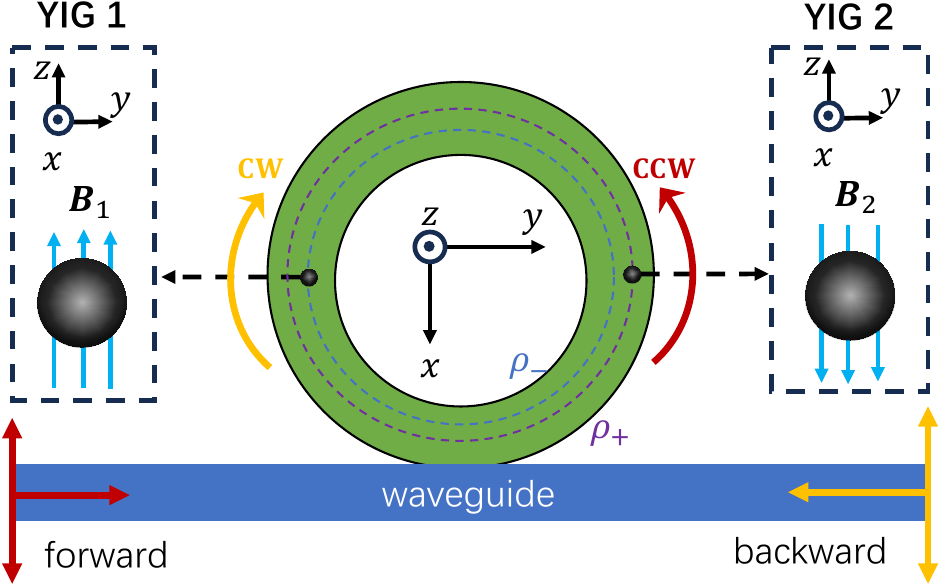}
\caption{The schematic diagram illustrates the cavity-magnetic subsystem, where the two YIG spheres exhibit distinct characteristics \cite{PhysRevB.103.104432}. The magnetic field of the cavity aligns with the $x$-axis, and both YIG spheres are coupled to it. The dashed lines indicate the special radial positions $\rho_{-}$ and $\rho_{+}$.}
\label{fig_01}
\end{figure}
We consider a hybrid magnon-cavity system consisting of two YIG spheres placed inside a torus-shaped microwave cavity (see Fig. \ref{fig_01}). The microwave cavity holds a pair of degenerate counter-propagating transverse electric (TE) microwave modes, which couple to a waveguide through which a specific circulating mode is driven. The Hamiltonian for microwave cavity modes reads $\hat{\mathcal{H}}_{c}=\omega_{c}(\hat{c}_{\ccw}^{\dag}\hat{c}_{\ccw}+\hat{c}_{\cw}^{\dag}\hat{c}_{\cw})$, where $\hat{c}_{\ccw}$ and $\hat{c}_{\cw}$ ($\hat{c}_{\ccw}^{\dag}$ and $\hat{c}_{\cw}^{\dag}$) denote the annihilation (creation) operators of counter-clockwise (CCW) and clockwise (CW) circulating microwave modes. There exist two distinct radial positions, denoted as $\rho_{\pm}$, within the ring cavity where the magnetic field exhibits circular polarization. For the CW mode, the magnetic field at $\rho_{-}$ exhibits left-handed polarization, whereas at $\rho_{+}$, it shows right-handed polarization. In contrast, for the CCW mode, the polarization characteristics are reversed \cite{PhysRevB.102.064416,PhysRevApplied.19.014030}. Two YIG spheres in the cavity are magnetized to saturation by external bias magnetic fields aligned along the $\textbf{z}$ and $-\textbf{z}$ direction. Then the Hamiltonian to describe YIG spheres reads $\hat{\mathcal{H}}_{m}=-\gamma B_{1}\hat{S}_{1}^{z}+\gamma B_{2}\hat{S}_{2}^{z}$, where $B_{j}$ denotes the magnitude of the external magnetic fields $\textbf{B}_{j}$ acting on the $j$-th YIG sphere ($j=1,2$), and $\gamma$ is the gyromagnetic ratio. The YIG spheres are coupled to microwave mode via the magnetic-dipole interaction. In this work, we consider the case that the CCW circulating microwave mode is polarized in the torus-shaped cavity. Then the Hamiltonian of chiral magnon-cavity coupling reads,
\begin{equation}\label{eq_01}
\begin{split}
\hat{\mathcal{H}}_{cm}=&\omega_{c}\hat{c}_{\ccw}^{\dagger}\hat{c}_{\ccw}-\gamma B_{1}\hat{S}_{1}^{z}
+\gamma B_{2}\hat{S}_{2}^{z}\\
&+\gamma\sqrt{\frac{\mu_{0}\omega_{c}}{4V_{c}}}(\hat{S}_{1}^{+}\hat{c}_{\ccw}+\hat{S}_{1}^{-}\hat{c}_{\ccw}^{\dagger})\\
&+\gamma\sqrt{\frac{\mu_{0}\omega_{c}}{4V_{c}}}(\hat{S}_{2}^{+}\hat{c}_{\ccw}+\hat{S}_{2}^{-}\hat{c}_{\ccw}^{\dagger}),
\end{split}
\end{equation}
where $\hat{S}_{j}^{\pm}=\hat{S}_{j}^{x}\pm i\hat{S}_{j}^{y}$ are the collective raising and lowering operators for $j$-th YIG sphere. Given that two YIG spheres with distinct chirality are simultaneously coupled to the CCW mode at different radial positions, the CW mode can consequently achieve effective decoupling from the system. In particular, we consider the magnon mode to be the Kittle mode, which corresponds to the uniform procession of all spins in the YIG sphere.
Bosonic collective excitations in a macroscopic spin ensemble can be introduced via the Holstein-Primakoff (HP) transformation \cite{PhysRev.58.1098}, which maps spin operators to bosonic operators 
\begin{equation}\label{eq_02}
\begin{split}
\begin{aligned}
\hat{S}_{1}^{+}&=\hat{h}(S_{1},\hat{n}_{1})\hat{m}_{1},\\
\hat{S}_{1}^{-}&=\hat{m}_{1}^{\dagger}\hat{h}(S_{1},\hat{n}_{1}),\\
\hat{S}_{1}^{z}&=S_{1}-\hat{m}_{1}^{\dagger}\hat{m}_{1},\\
\end{aligned}
\quad
\begin{aligned}
\hat{S}_{2}^{+}&=\hat{m}_{2}^{\dagger}\hat{h}(S_{2},\hat{n}_{2}),\\
\hat{S}_{2}^{-}&=\hat{h}(S_{2},\hat{n}_{2})\hat{m}_{2},\\
\hat{S}_{2}^{z}&=-S_{2}+\hat{m}_{2}^{\dagger}\hat{m}_{2},
\end{aligned}
\end{split}
\end{equation}
where $\hat{h}(S_{j},\hat{n}_{j})=\sqrt{2S_{j}(1-\hat{n}_{j}/2S_{j})}$ with $\hat{n}_{j}=\hat{m}_{j}^{\dag}\hat{m}_{j}$, $\hat{m}_{j}$ ($\hat{m}_{j}^{\dagger}$) are the bosonic annihilation (creation) operators that describe the Kittel modes of the respective YIG spheres, and $[\hat{m}_{i},\hat{m}_{j}]=0$, $[\hat{m}_{i}^{\dagger},\hat{m}_{j}^{\dagger}]=0$, $[\hat{m}_{i},\hat{m}_{j}^{\dagger}]=\delta_{ij}$. $S_j$ represents the quantum number of the spin angular momentum corresponding to the $j$-th YIG sphere. It is assumed that for the low-lying excitations, and the condition $\langle\hat{n}_j\rangle\ll 2S_j$ holds. Based on this assumption, $\hat{h}(S_j,\hat{n}_j)$ can be approximated as $\sqrt{2S_j}$. As a result, the operators in Eq. (\ref{eq_02}) can be approximated as follows
\begin{equation}\label{eq_03}
\begin{split}
\begin{aligned}
\hat{S}_{1}^{+}&\approx\sqrt{2S_{1}}\hat{m}_{1},\\
\hat{S}_{1}^{-}&\approx\sqrt{2S_{1}}\hat{m}_{1}^{\dagger},\\
\hat{S}_{1}^{z}&=S_{1}-\hat{m}_{1}^{\dagger}\hat{m}_{1},\\
\end{aligned}
\quad
\begin{aligned}
\hat{S}_{2}^{+}&\approx\sqrt{2S_{2}}\hat{m}_{2}^{\dagger},\\
\hat{S}_{2}^{-}&\approx\sqrt{2S_{2}}\hat{m}_{2},\\
\hat{S}_{2}^{z}&=-S_{2}+\hat{m}_{2}^{\dagger}\hat{m}_{2}.
\end{aligned}
\end{split}
\end{equation}

Then the Hamiltonian in Eq. (\ref{eq_01}) can be recast as
\begin{equation}\label{eq_04}
\begin{split}
\hat{\mathcal{H}}_{cm}=&\omega_{c}\hat{c}_{\ccw}^{\dagger}\hat{c}_{\ccw}+\omega_{1}\hat{m}_{1}^{\dagger}\hat{m}_{1}+\omega_{2}\hat{m}_{2}^{\dagger}\hat{m}_{2}\\
&+g_{1}^{\prime}(\hat{c}_{\ccw}^{\dagger}\hat{m}_{1}^{\dagger}+\hat{c}_{\ccw}\hat{m}_{1}) +g_{2}(\hat{c}_{\ccw}^{\dagger}\hat{m}_{2}+\hat{c}_{\ccw}^{\dagger}\hat{m}_{2}),
\end{split}
\end{equation}
where the parameters $\omega_{j}=\gamma B_{j}$, $g_{1}^{\prime}=\gamma\sqrt{\mu_{0}S_{1}\omega_{c}/2V_{c}}$ and $g_{2}=\gamma\sqrt{\mu_{0}S_{2}\omega_{c}/2V_{c}}$.

\begin{figure}
\centering
\includegraphics[width=0.44\textwidth]{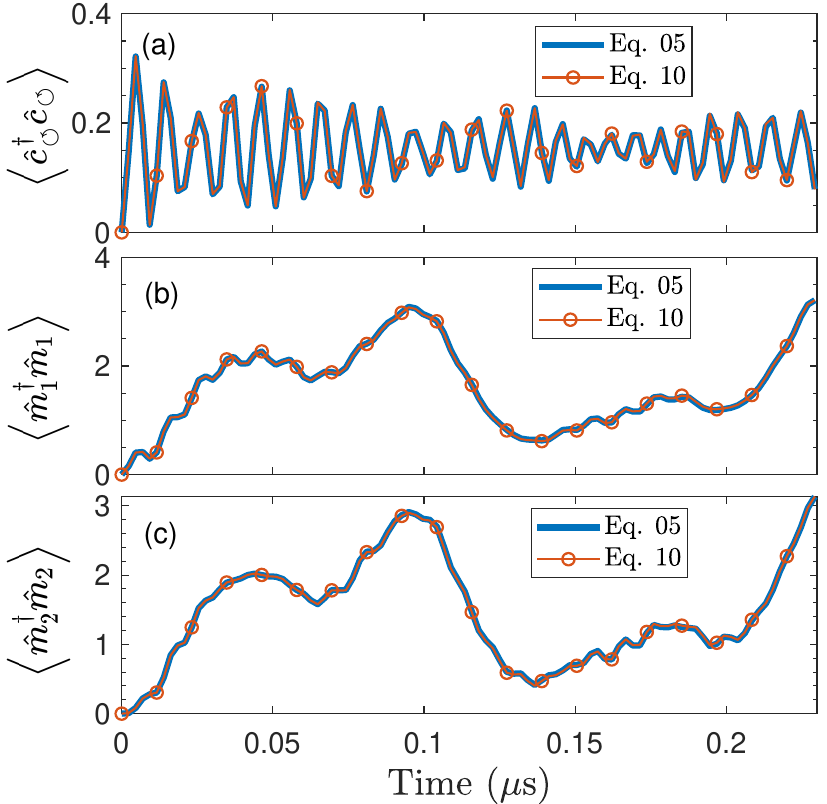}
\caption{The time-independent Hamiltonian Eq. (\ref{eq_10}) accurately captures the dynamic evolution of the system. The solid blue line and the solid red line with hollow circular markers represent the population of the cavity field and the two magnons, as described by the full Hamiltonian Eq. (\ref{eq_05}) and the time-independent Hamiltonian Eq. (\ref{eq_10}), respectively. Other parameters are as follows: $\omega_c/2\pi=10\text{ GHz}$, $\omega_1/2\pi=10.1\text{ GHz}$, $\omega_2/2\pi=9.9\text{ GHz}$, $\omega_d/2\pi=20\text{ GHz}$, $\Omega_d/2\pi=36\text{ GHz}$, $g_1^{\prime}/2\pi=50\text{ MHz}$, $g_2/2\pi=30\text{ MHz}$. Assume that the initial state of the system is prepared in the vacuum state $\left|0\right\rangle_{\hat{c}_{\ccw}}\otimes\left|0\right\rangle_{\hat{m}_{1}}\otimes\left|0\right\rangle_{\hat{m}_{2}}$, e.g., at a low bath temperature of tens of mK.}
\label{fig_02}
\end{figure}

It becomes clear that the coupling between the cavity field and the left YIG sphere (i.e., magnon $\hat{m}_{1}$) exhibits an anti-rotational characteristic. To facilitate subsequent analysis in Eq. (\ref{eq_04}), we applied a periodic modulation to magnon $\hat{m}_{1}$, expressed as $\hat{\mathcal{H}}_{d}(t)=\Omega_{d}\cos{(\omega_{d}t)}\hat{m}_{1}^{\dagger}\hat{m}_{1}$, where $\Omega$ denotes the driving strength and $\omega_{d}$ represents the driving frequency. Consequently, the total Hamiltonian of the system can be expressed as: 
\begin{equation}\label{eq_05}
\begin{split}
\hat{\mathcal{H}}_{T}(t)=&\hat{\mathcal{H}}_{cm} + \hat{\mathcal{H}}_{d}(t)\\
=&\omega_{c}\hat{c}_{\ccw}^{\dagger}\hat{c}_{\ccw}+\left[\omega_{1}+\Omega_{d}\cos{(\omega_{d}t)}\right]\hat{m}_{1}^{\dagger}\hat{m}_{1}\\
&+\omega_{2}\hat{m}_{2}^{\dagger}\hat{m}_{2}+g_{1}^{\prime}(\hat{c}_{\ccw}^{\dagger}\hat{m}_{1}^{\dagger}+\hat{c}_{\ccw}\hat{m}_{1})\\
& +g_{2}(\hat{c}_{\ccw}^{\dagger}\hat{m}_{2}+\hat{c}_{\ccw}\hat{m}_{2}^{\dagger}).
\end{split}
\end{equation}
The Hamiltonian in Eq. (\ref{eq_05}) can be transformed into an interaction picture $\hat{\mathcal{H}}_{T}^{(1)}(t)= U_{1}^{\dag}(t)[\hat{\mathcal{H}}_{T}(t)-i\partial_{t}]U_{1}(t)$. This can be accomplished by choosing
\begin{equation}\label{eq_06}
\begin{split}
\hat{U}_{1}(t)=&\exp\left[-i\left( \omega_{c}\hat{c}_{\ccw}^{\dagger}\hat{c}_{\ccw}+\omega_{1}\hat{m}_{1}^{\dagger}\hat{m}_{1} + \omega_{2}\hat{m}_{2}^{\dagger}\hat{m}_{2} \right)t \right. \\
& \left. - i\xi\sin(\omega_{d}t)\hat{m}_{1}^{\dagger}\hat{m}_{1}\right],
\end{split}
\end{equation}
where $\xi=\Omega_{d}/\omega_{d}$. After applying $\hat{U}_{1}(t)$ the transformed Hamiltonian in Eq. (\ref{eq_05}) can be rewritten as follows 
\begin{equation}\label{eq_07}
\begin{split}
\hat{\mathcal{H}}_{T}^{(1)}(t)
=&g_{1}^{\prime}(\hat{c}_{\ccw}^{\dagger}\hat{m}_{1}^{\dagger}{\rm e}^{i(\omega_{c}t+\omega_{1}t+\xi\sin{(\omega_{d}t)})}\\
&+\hat{c}_{\ccw}\hat{m}_{1}{\rm e}^{-i(\omega_{c}t+\omega_{1}t+\xi\sin{(\omega_{d}t)})})\\
&+g_{2}(\hat{c}_{\ccw}^{\dagger}\hat{m}_{2}{\rm e}^{i(\omega_{c}-\omega_{2})t}+\hat{c}_{\ccw}\hat{m}_{2}^{\dagger}{\rm e}^{-i(\omega_{c}-\omega_{2})t}).
\end{split}
\end{equation}
By employing the Jacobi-Anger expansion $\exp\left[-i\xi\sin{(\omega_{d}t)}\right]=\sum_{f=-\infty}^{+\infty} J_{f}(\xi) \exp(-if\omega_{d}t)$ with $J_{f}(\xi)$ being the $f$-th order Bessel function of the first kind, the Hamiltonian in Eq. (\ref{eq_07}) can be recast as
\begin{equation}\label{eq_08}
\begin{split}
\hat{\mathcal{H}}_{T}^{(2)}(t)
=&-g_{1}(\hat{c}_{\ccw}^{\dagger}\hat{m}_{1}^{\dagger}{\rm e}^{i\Delta_{1}t}+\hat{c}_{\ccw}\hat{m}_{1}{\rm e}^{-i\Delta_{1}t})\\
&+g_{2}(\hat{c}_{\ccw}^{\dagger}\hat{m}_{2}{\rm e}^{i\Delta_{2}t}+\hat{c}_{\ccw}\hat{m}_{2}^{\dagger}{\rm e}^{-i\Delta_{2}t})\\
&+g_{1}^{\prime}\sum_{f\ne-1}J_{f}(\xi)(\hat{c}_{\ccw}^{\dagger}\hat{m}_{1}^{\dagger}{\rm e}^{i\delta_{f}t}+\hat{c}_{\ccw}\hat{m}_{1}{\rm e}^{-i\delta_{f}t}),
\end{split}
\end{equation}
where $\Delta_{1}=\omega_{c}+\omega_{1}-\omega_{d}$, $\delta_{f}=\omega_{c}+\omega_{1}+f\omega_{d}$, $\Delta_{2}=\omega_{c}-\omega_{2}$ and $g_{1}=-g_{1}^{\prime}J_{-1}(\xi)$.
Here, we assume that $\left|\delta_{f}\right|\gg g_{1}^{\prime}J_{f}(\xi)$ ($f\ne-1$), allowing us to neglect the high-frequency terms, specifically the non-resonant contributions. Consequently, we obtain the following simplified Hamiltonian
\begin{equation}\label{eq_09}
\begin{split}
\hat{\mathcal{H}}_{T}^{(3)}=&-g_{1}(\hat{c}_{\ccw}^{\dagger}\hat{m}_{1}^{\dagger}e^{i\Delta_{1}t}+\hat{c}_{\ccw}\hat{m}_{1}e^{-i\Delta_{1}t})\\
&+g_{2}(\hat{c}_{\ccw}^{\dagger}\hat{m}_{2}e^{i\Delta_{2}t}+\hat{c}_{\ccw}\hat{m}_{2}^{\dagger}e^{-i\Delta_{2}t}),
\end{split}
\end{equation}
where the coupling strength between the cavity field and the magnon $\hat{m}_{1}$ has been modulated by the periodic driving. Moving to the rotating frame defined by $U_{2}(t)=\exp{[i\omega_{c}\hat{c}_{\ccw}^{\dagger}\hat{c}_{\ccw}t+i(\omega_{1}-\omega_{d})\hat{m}_{1}^{\dagger}\hat{m}_{1}t+i\omega_{2}\hat{m}_{2}^{\dagger}\hat{m}_{2}t]}$, we obtain the transformed Hamiltonian $\hat{\mathcal{H}}_{T}^{(4)}=\hat{U}_{2}(t)^{\dag}[\hat{\mathcal{H}}_{T}^{(3)}(t)-i\partial_{t}]\hat{U}_{2}(t)$ as follows
\begin{equation}\label{eq_10}
\begin{split}
\hat{\mathcal{H}}_{T}^{(4)}=&\hat{\mathcal{H}}_{0}+\hat{\mathcal{H}}_{\rm int},
\end{split}
\end{equation}
where $\hat{\mathcal{H}}_{0}$ is the free part of the system and $\hat{\mathcal{H}}_{\rm int}$ represents the interaction term. Concretely,
\begin{equation}\label{eq_11}
\begin{split}
\hat{\mathcal{H}}_{0}=&\omega_{c}\hat{c}_{\ccw}^{\dagger}\hat{c}_{\ccw}+\varpi\hat{m}_{1}^{\dagger}\hat{m}_{1}+\omega_{2}\hat{m}_{2}^{\dagger}\hat{m}_{2},\\
\hat{\mathcal{H}}_{\rm int}=&-g_{1}(\hat{c}_{\ccw}^{\dagger}\hat{m}_{1}^{\dagger}+\hat{c}_{\ccw}\hat{m}_{1})+g_{2}(\hat{c}_{\ccw}^{\dagger}\hat{m}_{2}+\hat{c}_{\ccw}\hat{m}_{2}^{\dagger}),
\end{split}
\end{equation}
where $\varpi=\omega_{1}-\omega_{d}$. To validate the assumption regarding the negligibility of higher-order perturbation terms in the theoretical model, this study systematically compares the populations of the CCW cavity mode and the two magnon modes between the simplified time-independent Hamiltonian Eq. (\ref{eq_10}) and the full Hamiltonian Eq. (\ref{eq_05}) under parameter configurations satisfying steady-state conditions (selected based on steady-state criterion). As illustrated in Fig. \ref{fig_02}, the temporal evolution of the cavity mode and the two magnon modes population exhibits only minor discrepancies and ultimately converges to a stable state. This observation confirms that neglecting higher-order terms in the original theoretical derivation is rigorously justified. In the following section, we analyze the practical applications of the simplified time-independent Hamiltonian Eq. (\ref{eq_10}) under various constraints. 

\section{Magnon-magnon entanglement and squeezing}
\label{Sec.III}
In this section, we investigate the entanglement between two magnon modes. To analyze this scenario comprehensively, we model the system dynamics by incorporating dissipative processes through the Heisenberg-Langevin equations. The quantum Langevin equation for the dynamical evolution of the system can be written as
\begin{equation}\label{eq_12}
\begin{split}
\frac{d}{dt}\hat{c}_{\ccw}&=-(i\omega_{c}+\frac{\kappa}{2})\hat{c}_{\ccw}+ig_{1}\hat{m}_{1}^{\dagger }-ig_{2}\hat{m}_{2}+\sqrt{\kappa}\hat{c}_{\ccw,in},\\
\frac{d}{dt}\hat{m}_{1}&=-(i\varpi+\frac{\gamma_{1}}{2})\hat{m}_{1}+ig_{1} \hat{c}_{\ccw}^{\dagger }+\sqrt{\gamma_{1}}\hat{m}_{1,in},\\
\frac{d}{dt}\hat{m}_{2}&=-(i\omega_{2}+\frac{\gamma_{2}}{2})\hat{m}_{2}-ig_{2}\hat{c}_{\ccw}+\sqrt{\gamma_{2}}\hat{m}_{2,in},\\
\frac{d}{dt}\hat{c}_{\ccw}^{\dagger}&=(i\omega_{c}-\frac{\kappa}{2})\hat{c}_{\ccw}^{\dagger}+ig_{1}\hat{m}_{1}+ig_{2}\hat{m}_{2}^{\dagger}+\sqrt{\kappa}\hat{c}_{\ccw,in}^{\dagger},\\
\frac{d}{dt}\hat{m}_{1}^{\dagger}&=(i\varpi-\frac{\gamma_{1}}{2})\hat{m}_{1}^{\dagger}-ig_{1}\hat{c}_{\ccw}+\sqrt{\gamma_{1}}\hat{m}_{1,in}^{\dagger},\\
\frac{d}{dt}\hat{m}_{2}^{\dagger}&=(i\omega_{2}-\frac{\gamma_{2}}{2})\hat{m}_{2}^{\dagger}-ig_{2}\hat{c}_{\ccw}^{\dagger}+\sqrt{\gamma_{2}}\hat{m}_{2,in}^{\dagger}.
\end{split}
\end{equation}
The notations $\hat{c}_{\ccw,in}$ and $m_{j,in}$ represent the input noise operator with zero mean average, which can be characterized by the correlation function \cite{Gardiner2004QuantumNA}: $\langle\hat{c}_{\ccw,in}(t)\hat{c}_{\ccw,in}^{\dagger}(t^{\prime})\rangle=\delta(t-t^{\prime})$ and $\langle \hat{m}_{j,in}(t)\hat{m}_{j,in}^{\dagger}(t^{\prime})\rangle=\delta(t-t^{\prime})$.

By further defining the quadrature components of the Kittle mode and the cavity mode as $\hat{X}_{O}=(\hat{O}^{\dagger}+\hat{O})/\sqrt{2}$ and $\hat{Y}_{O}=i(\hat{O}^{\dagger}-\hat{O})/\sqrt{2}$, and the associated input noise operators as $\hat{X}_{O}^{in}=(\hat{O}_{in}^{\dagger}+\hat{O}_{in})/\sqrt{2}$ and $\hat{Y}_{O}^{in}=i(\hat{O}_{in}^{\dagger}-\hat{O}_{in})/\sqrt{2}$ where $\hat{O}=\{\hat{c}_{\ccw}, \hat{m}_{1}, \hat{m}_{2}\}$. Hence, the dynamics in Eq. (\ref{eq_12}) can be equivalently expressed in the following matrix form:
\begin{equation}\label{eq_13}
\frac{d}{dt}\textbf{R}(t)=\textbf{A}\textbf{R}(t)+\textbf{N},
\end{equation}
where $\textbf{R}(t)=[\hat{X}_{c},\hat{Y}_{c},\hat{X}_{m_1},\hat{Y}_{m_1},\hat{X}_{m_2},\hat{Y}_{m_2}]^{T}$ is the state vector of the system, and $\textbf{N}=[\sqrt{\kappa}\hat{X}_{c}^{in}, \sqrt{\kappa}\hat{Y}_{c}^{in},\sqrt{\gamma_{1}}\hat{X}_{m_1}^{in}, \sqrt{\gamma_{1}}\hat{Y}_{m_1}^{in}, \sqrt{\gamma_{2}}\hat{X}_{m_2}^{in}, \sqrt{\gamma_{2}}\hat{Y}_{m_2}^{in}]^{T}$ is the input noise vector. The coefficient matrix $\textbf{A}$ is given by 
\begin{equation}\label{eq_14}
\begin{split}
\textbf{A}=\begin{pmatrix}
-\frac{\kappa}{2}&\omega_{c}&0&g_{1}&0&g_{2}\\
-\omega_{c}&-\frac{\kappa}{2}&g_{1}&0&-g_{2}&0\\
0&g_{1}&-\frac{\gamma_{1}}{2}&\varpi&0&0\\
g_{1}&0&-\varpi&-\frac{\gamma_{1}}{2}&0&0\\
0&g_{2}&0&0&-\frac{\gamma_{2}}{2}&\omega_{2}\\
-g_{2}&0&0&0&-\omega_{2}&-\frac{\gamma_{2}}{2}
\end{pmatrix}.
\end{split}
\end{equation}
Since we are using the linear quantum Langevin equations, the Gaussian properties of the input states are maintained throughout the evolution of the systems. Consequently, the quantum fluctuations manifest as a continuously evolving three-mode Gaussian state \cite{PhysRevResearch.1.023021}, which can be completely characterized by a $6\times6$ covariance matrix (CM) $\bm{\sigma}(t)$ \cite{PhysRevLett.98.030405}, defined as $\sigma_{mn}(t)=\langle \textbf{R}_{m}\textbf{R}_{n}+\textbf{R}_{n}\textbf{R}_{m}\rangle/2$ ($m,n=1,2,...,6$). We can substitute the above form into Eq. (\ref{eq_13}), and the dynamical evolution of $\bm{\sigma}(t)$ is given by \cite{PhysRevA.103.033508}
\begin{equation}\label{eq_15}
\begin{split}
\frac{d}{dt}\bm{\sigma}(t)=\textbf{A}\bm{\sigma}(t)+\bm{\sigma}(t )\textbf{A}^{T}+\textbf{D},
\end{split}
\end{equation}
where $\textbf{A}^{T}$ denotes the transpose of the matrix $\textbf{A}$, and $\textbf{D}$ represents the diffusion matrix, given by $\textbf{D}=\rm diag\left[\kappa/2,\kappa/2,\gamma_{1}/2,\gamma_{1}/2,\gamma_{2}/2,\gamma_{2}/2\right]$. Once the system's covariance matrix $\bm{\sigma}(t)$ is obtained, the entanglement between the cavity and the two magnons can be quantified using the logarithmic negativity. The covariance matrix $\bm{\sigma}(t)$ can be expressed in the form of block matrix:
\begin{equation}\label{eq_16}
\begin{split}
\bm{\sigma}(t)=\begin{pmatrix}
\bm{\Phi}_{\hat{c}_{\ccw}}&\bm{\Phi}_{\hat{c}_{\ccw}\hat{m}_{1}}&\bm{\Phi}_{\hat{c}_{\ccw}\hat{m}_{2}}\\
\bm{\Phi}_{\hat{c}_{\ccw}\hat{m}_{1}}^{T}&\bm{\Phi}_{\hat{m}_{1}}&\bm{\Phi}_{\hat{m}_{1}\hat{m}_{2}}\\
\bm{\Phi}_{\hat{c}_{\ccw}\hat{m}_{2}}^{T}&\bm{\Phi}_{\hat{m}_{1}\hat{m}_{1}}^{T}&\bm{\Phi}_{\hat{m}_{2}}
\end{pmatrix},
\end{split}
\end{equation}
where each block is a $2\times2$ matrix. Here the diagonal block $\bm{\Phi}_{\mu}$ represents the system $\mu$$(\mu=\hat{c}_{\ccw},\hat{m}_{1},\hat{m}_{2})$, and $\bm{\Phi}_{\mu,\nu}$ is the sub-block matrix related to the correlation of systems $\mu$ and $\nu$. To explore the entanglement between any two subsystems of the cavity and two magnons, we extract the relevant $4\times4$ reduced covariance matrix $\bm{\sigma}^{\prime}(t)$ from the full covariance matrix $\bm{\sigma}(t)$. By considering the covariance matrix
\begin{equation}\label{eq_17}
\begin{split}
\bm{\sigma}_{\mu\nu}^{\prime}(t)=\begin{pmatrix}
\bm{\Phi}_{\mu}&\bm{\Phi}_{\mu\nu}\\
\bm{\Phi}_{\mu\nu}^{T}&\bm{\Phi}_{\nu}
\end{pmatrix}.
\end{split}
\end{equation}
In the CV case, we can derive the entanglement degree $E_{N}$ as \cite{PhysRevA.70.022318,Adesso_2007,PhysRevB.101.014419}
\begin{equation}\label{eq_18}
\begin{split}
E^{N}_{\mu\nu}=\rm Max[0,-ln(2\eta^{-}_{\mu\nu})],
\end{split}
\end{equation}
where
\begin{equation}\label{eq_19}
\begin{split}
\eta^{-}_{\mu\nu}&=\sqrt{\left(W_{\mu\nu}-\sqrt{W_{\mu\nu}^{2}-4{\rm Det}\left[\bm{\sigma}^{\prime}_{\mu\nu}\right]}\right)\big/2},\\
W_{\mu\nu}&={\rm Det}\left[\bm{\Phi}_{\mu}\right]+ {\rm Det}\left[\bm{\Phi}_{\nu}\right]-2{\rm Det}\left[\bm{\Phi}_{\mu\nu}\right],
\end{split}
\end{equation}
and the notation $\operatorname{Det}[\bullet]$ represents the determinant of a matrix.

\begin{figure}
\centering
\includegraphics[width=0.48\textwidth]{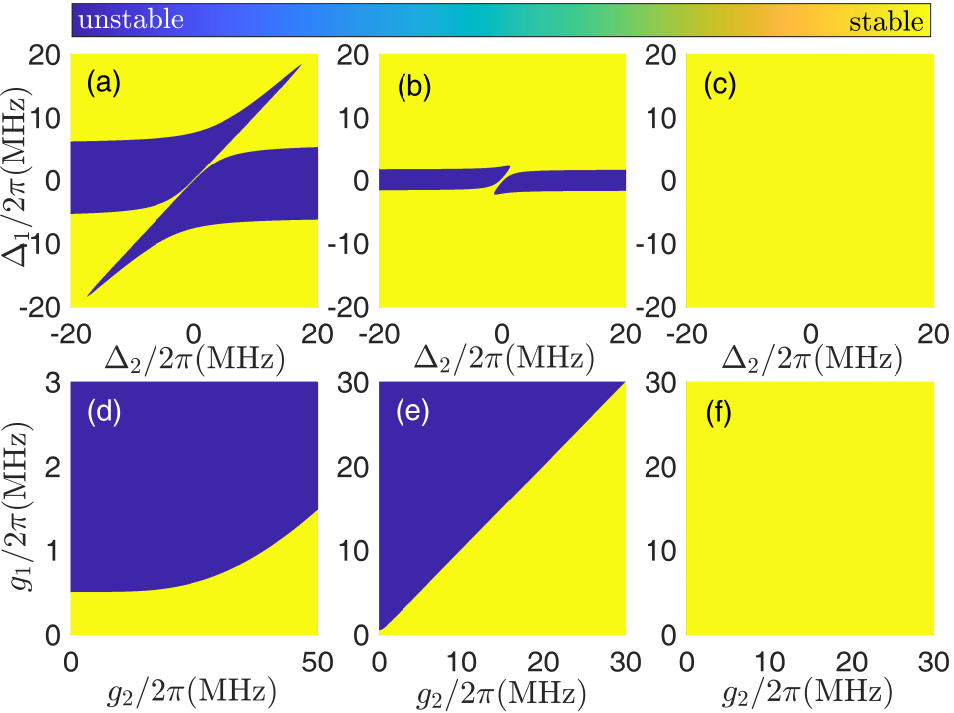}
\caption{The stability characteristics of the cavity-magnon system are systematically investigated under the following conditions: (a) $g_{1}=g_{2}=3\text{ MHz}$, (b) $g_{1}=g_{2}=1\text{ MHz}$, and (c) $g_{1}=g_{2}=0.5\text{ MHz}$, with respect to the detuning parameters $\Delta_{1}$ and $\Delta_{2}$. Furthermore, the stability of the cavity-magnetic resonance subsystem is analyzed as a function of the coupling strengths $g_{1}$ and $g_{2}$ for two specific cases: (d) $\Delta_{1}=0$ and $\Delta_{2}=0.9\text{ GHz}$, (e) $\Delta_{1}=\Delta_{2}=0$ and (f) $\Delta_{1}=\Delta_{2}=0.9\text{ GHz}$. The system parameters are fixed at $\omega_{c}/2\pi=10\text{ GHz}, \omega_{d}/2\pi=20\text{ GHz}, \kappa/2\pi= \gamma_{1}/2\pi=\gamma_{2}/2\pi=1\text{ MHz}$. In the presented results, stable and unstable regions are represented by yellow and blue areas, respectively.}
\label{fig_03}
\end{figure}

\subsection{Stable condition}
\label{Sec.IIIa}

The stability of a system is a prerequisite for investigating quantum coherence. The system becomes stable and attains its steady state only when the real parts of $\textbf{A}$ are negative. The corresponding necessary conditions for stability can be derived using the Routh-Hurwitz criterion \cite{PhysRevA.35.5288}. When the system attains a steady state, the dynamical Eq. (\ref{eq_15}) governing the covariance matrix $\bm{\sigma}$ can be simplified to the following Lyapunov equation:
\begin{equation}\label{eq_20}
\textbf{A}\bm{\sigma}+\bm{\sigma}\textbf{A}^{T}=-\textbf{D}.
\end{equation}

To determine the parameter regime of stable steady states, the characteristic equation $|\lambda\textbf{I}- \textbf{A}|=0$ is analytically solved (see Appendix \ref{app_A} for details). However, due to the complexity of our system, a direct calculation of the steady-state parameter regime is analytically intractable. Therefore, we use numerical stability diagrams for a more intuitive analysis. Based on the results shown in Fig. \ref{fig_03}, we analyzed the stability of the system under different detunings and coupling parameters. Preliminary studies indicate that, as illustrated in Figs. \ref{fig_03}(a), \ref{fig_03}(b), and \ref{fig_03}(c), when the coupling strengths $g_{1}$ and $g_{2}$ are both less than or equal to $0.5\text{ MHz}$ (i.e., $g_{1}=g_{2}\leq 0.5\text{ MHz}$), the system remains stable for all values of $\Delta_{1}$ and $\Delta_{2}$. Furthermore, we found that when $\Delta_{1}=0$ and $\Delta_{2}\neq 0$, the system is prone to instability. Consequently, we plotted the stability as a function of the coupling strengths $g_{1}$ and $g_{2}$ for the case where $\Delta_{1}=0$ and $\Delta_{2}=0.9\text{ GHz}$, as shown in Fig. \ref{fig_03}(d). From the analysis of the figure, we observed that stability occurs when $g_{2} > g_{1}$. Moreover, as shown in Fig. \ref{fig_03}(e), the system remains stable under the condition of $g_{2}\ge g_{1}$ when $\Delta_{1}=\Delta_{2}=0$. Finally, Fig. \ref{fig_03}(f) demonstrates that under the large detuning condition ($\Delta_{1},\Delta_{2}\gg g_{j}$), the system maintains stability for any values of $g_{1}$ and $g_{2}$.


\subsection{Sideband cooling technique and magnonic two-mode squeezed states}
\label{Sec.IIIb}

To determine the maximum entanglement achievable in this system, we varied the detunings $\Delta_{1}$ and $\Delta_{2}$, while keeping all other experimental parameters constant. The entanglement between the cavity and the magnon $\hat{m}_{1}$, the cavity and magnon $\hat{m}_{2}$, as well as between the two magnons, is illustrated in Fig. \ref{fig_02}. The parameter range we selected ensures that the coefficient matrix \textbf{A} in Eq. (\ref{eq_13}) satisfies the Routh-Hurwitz stability criterion.

\begin{figure}[htbp]
\centering
\includegraphics[width=0.48\textwidth]{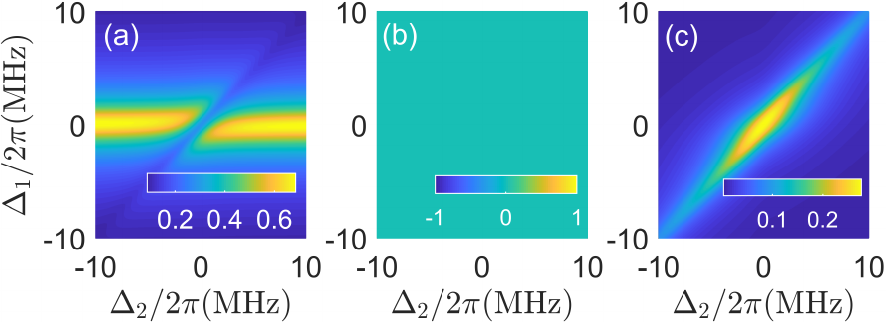}
\caption{The density plot illustrates the entanglement among the bipartite entanglements: (a) $E_{\hat{c}_{\ccw}\hat{m}_{1}}$, (b) $E_{\hat{c}_{\ccw}\hat{m}_{2}}$, and (c) $E_{\hat{m}_{1}\hat{m}_{2}}$ as a function of the detunings $\Delta_{1}$ and $\Delta_{2}$. Other parameters are as follows: $\omega_{c}/2\pi=10\text{ GHz}, \omega_{d}/2\pi=20\text{ GHz}, g_{1}/2\pi=1\text{ MHz}, g_{2}/2\pi=0.5\text{ MHz}, \kappa/2\pi= \gamma_{1}/2\pi=\gamma_{2}/2\pi=1\text{ MHz}$. Assume that the initial state of the system is prepared in the ground state $\left|0\right\rangle_{\hat{c}_{\ccw}}\otimes\left|0\right\rangle_{\hat{m}_{1}}\otimes\left|0\right\rangle_{\hat{m}_{2}}$.}
\label{fig_04}
\end{figure}

Compared to Figs. \ref{fig_04}(a), \ref{fig_04}(b), and \ref{fig_04}(c), it is evident that when the detunings are equal, i.e., $\Delta_{1}=\Delta_{2}$, significant entanglement occurs between the two magnons. Furthermore, the entanglement between the two magnons reaches its maximum when $\Delta_{1}=\Delta_{2}=0$. To elucidate this phenomenon, we set $\Delta_{1}$ and $\Delta_{2}$ to zero in the Hamiltonian Eq. (\ref{eq_09}), allowing us to derive the following results:
\begin{equation}\label{eq_21}
\begin{split}
\hat{\mathcal{H}}=-g_{1}(\hat{c}_{\ccw}^{\dagger}\hat{m}_{1}^{\dagger}+\hat{c}_{\ccw}\hat{m}_{1})+g_{2}(\hat{c}_{\ccw}^{\dagger}\hat{m}_{2}+\hat{c}_{\ccw}\hat{m}_{2}^{\dagger}).
\end{split}
\end{equation}
This form of the Hamiltonian incorporates the fundamental interactions necessary to create entanglement between the modes $\hat{m}_{1}$ and $\hat{m}_{2}$. Initially, the parametric amplification interaction $-g_{1}(\hat{c}_{\ccw}^{\dagger}\hat{m}_{1}^{\dagger}+\hat{c}_{\ccw}\hat{m}_{1})$ creates entanglement between mode $\hat{c}_{\ccw}$ and mode $\hat{m}_{1}$. Subsequently, the beam splitter interaction $g_{2}(\hat{c}_{\ccw}^{\dagger}\hat{m}_{2}+\hat{c}_{\ccw}\hat{m}_{2}^{\dagger})$ facilitates the exchange of states between mode $\hat{c}$ and mode $\hat{m}_{2}$. Therefore, by using the cavity mode $\hat{c}_{\ccw}$ as an intermediary, the entanglement between the target modes $\hat{m}_{1}$ and $\hat{m}_{2}$ is effectively generated. To gain a deeper understanding of the physical mechanisms underlying entanglement generation, we introduce the nonlocal Bogoliubov mode operators for the two magnons $\hat{m}_{1}$ and $\hat{m}_{2}$:
\begin{equation}\label{eq_22}
\begin{split}
\hat{\alpha}=&\hat{m}_{1}\cosh{r}-\hat{m}_{2}^{\dagger}\sinh{r}=\hat{S}(r)\hat{m}_{1}\hat{S}^{\dagger}(r),\\
\hat{\beta}=&\hat{m}_{2}\cosh{r}-\hat{m}_{1}^{\dagger}\sinh{r}=\hat{S}(r)\hat{m}_{2}\hat{S}^{\dagger}(r),
\end{split}
\end{equation}
where $\hat{S}(r)=\exp{[r(\hat{m}_{1}\hat{m}_{2}-\hat{m}_{1}^{\dagger}\hat{m}_{2}^{\dagger})]}$ is the two-mode squeezing operator and $r=\arctanh{(g_{1}/g_{2})}$ is the squeezing amplitude. From the definitions, we observe that both modes $\hat{\alpha}$ and $\hat{\beta}$ are in the squeezed vacuum state given by $|r\rangle=\hat{S}(r)|0,0\rangle$, where $\hat{\alpha}|r\rangle=\hat{\beta}|r\rangle=0$. Therefore, if the Bogoliubov modes $\hat{\alpha}$ or $\hat{\beta}$ are consistently maintained in their ground states, strong quantum entanglement will be established between the magnon modes $\hat{m}_{1}$ and $\hat{m}_{2}$ \cite{PhysRevB.101.014419}. By substituting the defined Bogoliubov mode operators Eq. (\ref{eq_22}) into the resonant Hamiltonian Eq. (\ref{eq_21}), we obtain the following expression:
\begin{equation}\label{eq_23}
\begin{split}
\hat{\mathcal{H}}=J(\hat{c}_{\ccw}\hat{\beta}^{\dagger}+\hat{c}_{\ccw}^{\dagger}\hat{\beta}),
\end{split}
\end{equation}
where $J=\sqrt{g_{2}^{2}-g_{1}^{2}}$ represents the effective coupling strength between the cavity $\hat{c}_{\ccw}$ and the Bogoliubov mode $\hat{\beta}$. From Eq. (\ref{eq_23}), it is evident that mode $\hat{\alpha}$ is decoupled, and we refer to it as the dark mode. Through beam splitter interactions, the Bogoliubov mode $\hat{\beta}$ couples with the cavity mode $\hat{c}_{\ccw}$. Furthermore, we observe that the cavity mode $\hat{c}_{\ccw}$ can cool the Bogoliubov mode $\hat{\beta}$ and maintain it consistently in its ground state. This process transfers the thermal population that would otherwise accumulate in $\hat{\beta}$ into the cavity mode, followed by subsequent dissipation into the environment. 

To identify the squeezing parameters that maximize entanglement, we can use the Hamiltonian in Eq. (\ref{eq_21}) to construct the corresponding coefficient matrix $\textbf{A}_{1}$:
\begin{equation}\label{eq_24}
\begin{split}
\textbf{A}_{1}=\begin{pmatrix}
-\frac{\kappa}{2}&0&0&g_{1}&0&g_{2}\\
0&-\frac{\kappa}{2}&g_{1}&0&-g_{2}&0\\
0&g_{1}&-\frac{\gamma_{1}}{2}&0&0&0\\
g_{1}&0&0&-\frac{\gamma_{1}}{2}&0&0\\
0&g_{2}&0&0&-\frac{\gamma_{2}}{2}&0\\
-g_{2}&0&0&0&0&-\frac{\gamma_{2}}{2}
\end{pmatrix}.
\end{split}
\end{equation}
Under the special resonant condition $\Delta_{1}=\Delta_{2}=0$, the stability condition of the system can be derived as $g_{2}\ge g_{1}$, according to the Routh-Hurwitz criterion (see Sec. \ref{Sec.IIIa} for details).

\begin{figure}
\centering
\includegraphics[width=0.46\textwidth]{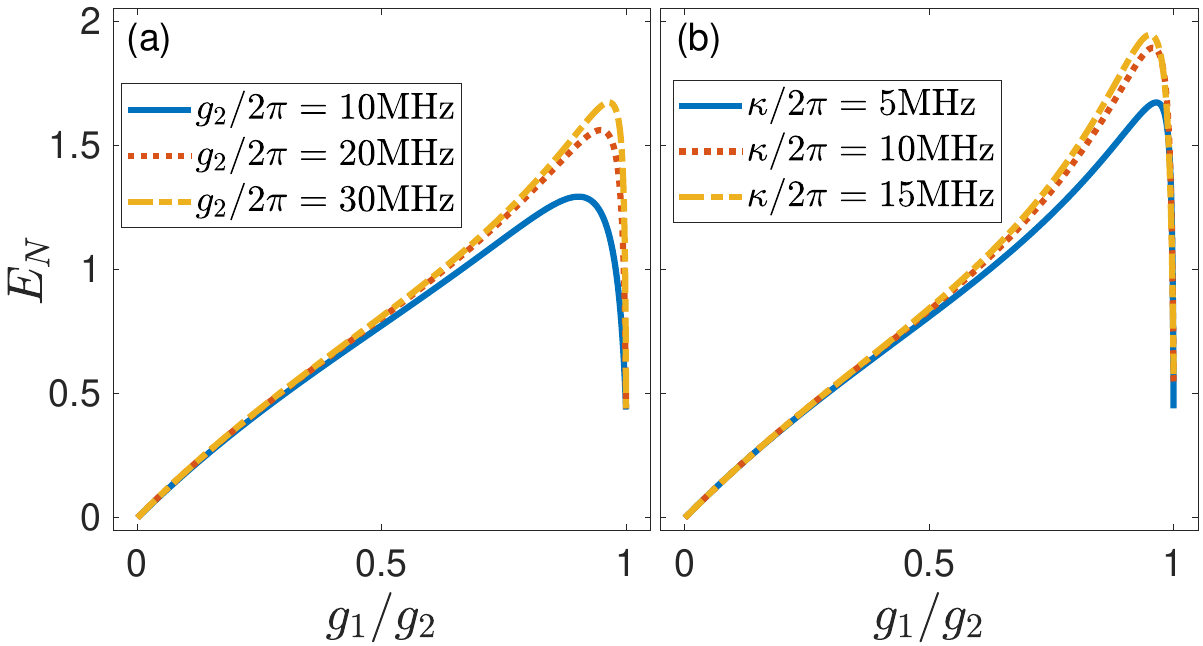}
\caption{Effect of the effective coupling sideband ratio $g_{1}/g_{2}$ on the entanglement between magnons. (a) Influence of different coupling strengths $g_{2}$ on entanglement, with $\kappa/2\pi=5\text{ MHz}$. (b) Influence of different cavity decay rates $\kappa$ on entanglement, with $g_{2}/2\pi=50\text{ MHz}$. Other parameters: $\gamma_{1}/2\pi=\gamma_{2}/2\pi=1\text{ MHz}$.}
\label{fig_05}
\end{figure}

Fig. \ref{fig_05}(a) illustrates the variation of the steady-state entanglement $E_{N}$ between the magnon modes as a function of the effective coupling sideband ratio $g_{1}/g_{2}$. The study reveals that the entanglement $E_{N}$ is a non-monotonic function of the ratio $g_{1}/g_{2}$. This phenomenon is attributed to the increase in $g_{1}/g_{2}$, which leads to a greater squeezing parameter $r=\arctanh{(g_{1}/g_{2})}$, thereby enhancing the entanglement between the magnon modes $\hat{m}_{1}$ and $\hat{m}_{2}$. Conversely, when considering the effective coupling strength $J=\sqrt{g_{2}^{2}-g_{1}^{2}}$, an increase in $g_{1}$ while holding $g_{2}$ fixed weakens $J$. This, in turn, inevitably suppresses the cooling capability of the Bogoliubov mode $\hat{\beta}$. Therefore, during the initial phase, as $g_{1}$ increases, the entanglement $E_{N}$ between the magnons strengthens. However, when the value of $g_{1}$, undergoes a continuous increase, the value of $J$ gradually approaches zero, resulting in a significant reduction in cooling capacity. Once the Bogoliubov mode $\hat{\beta}$ ceases to cool effectively, the entanglement $E_{N}$ between the magnons will rapidly decrease. Simultaneously, we observe that as $g_{2}$ increases, the optimal ratio $g_{1}/g_{2}$ gradually approaches 1. This phenomenon is based on the analysis of the effective coupling strength $J=\sqrt{g_{2}^{2}-g_{1}^{2}}=g_{2}\sqrt{1-(g_{1}/g_{2})^{2}}$. As $g_{2}$ increases, a larger $g_{1}/g_{2}$ ratio can still effectively ensure the cooling of mode $\hat{\beta}$.

Next, we investigate the impact of cavity field dissipation on the entanglement between two magnons. As shown in Fig. \ref{fig_05}(b), the entanglement between the two magnons increases significantly with increasing cavity field dissipation. Therefore, increasing the cavity mode attenuation rate $\kappa$ within a reasonable range can generate strong entanglement between the magnons. This is attributed to the fact that a higher cavity mode decay rate $\kappa$ enhances the cooling capability of the cavity mode $\hat{\beta}$, which, in turn, leads to stronger entanglement between the magnons.

In addition, we can introduce the effective dissipation $\Gamma=4J^{2}/\kappa$ \cite{PhysRevA.89.063805} of the new mode $\hat{\beta}$, to explain the phenomenon observed in Fig. \ref{fig_05}. To ensure that $\hat{\beta}$ remains in its ground state (i.e., the two-mode squeezed state), the dissipation rate must satisfy the condition $\Gamma\gg\gamma_{j}$. This is because if the dissipation $\gamma$ of the magnon modes $\hat{m}_{1}$  and $\hat{m}_{2}$ exceeds the effective dissipation of $\hat{\beta}$, the magnon modes $\hat{m}_{1}$ and $\hat{m}_{2}$ will rapidly relax to their vacuum states through their strong dissipation channels. This process would destroy the quantum coherence of the $\hat{\beta}$ mode, thereby preventing the system from maintaining $\hat{\beta}$ in its ground state. The experimental realization of this protocol requires dual constraints on system parameters. First, the effective coupling strength $J$ must satisfy $4J^{2}/\kappa\gg\gamma_{j}$. As shown in Fig. \ref{fig_05}(a), increasing the coupling strength $g_{2}$ significantly enhances the entanglement. However, when $J$ drops below a critical value, the entanglement is observed to decrease sharply. Secondly, the cavity dissipation rate $\kappa$ must remain low. As illustrated in Fig. \ref{fig_05} (b), properly increasing $\kappa$ can enhance the entanglement between the two magnons. However, the entanglement enhancement progressively diminishes with increasing $\kappa$, and an excessively large $\kappa$ ultimately degrades the entanglement. This phenomenon arises from the requirement to satisfy the condition $4J^{2}/\kappa\gg\gamma_{j}$, which imposes an upper bound on $\kappa$. Consequently, overly strong dissipation suppresses entanglement generation due to this critical condition. When both constraints are met, the effective dissipation rate $\Gamma$ dominates during the dynamical evolution of the system.

\subsection{Preparation of a two-mode magnonic squeezed state condition under large detuning}
\label{Sec.IIIc}
We perform a comparative analysis, as shown in Fig. \ref{fig_04}. In Figs. \ref{fig_04}(a), \ref{fig_04}(b), and \ref{fig_04}(c), we observe significant entanglement between the two magnons under conditions of large detuning, while no entanglement is detected between the magnons and the cavity field. Notably, when $\Delta_{1}\approx\Delta_{2}$, the entanglement between the two magnons is particularly pronounced. To better understand the interaction between the two magnons and obtain the effective  Hamiltonian in large detuning case (i.e., $g_{j}\ll\Delta_{j}$), we apply Schrieffer-Wolff perturbation theory (SWPT). By means of unitary transformation $\hat{U}=e^{i\hat{S}}$ with $\hat{S}=\frac{ig_{1}}{\Delta_{1}}(\hat{c}_{\ccw}^{\dagger}\hat{m}_{1}^{\dagger}-\hat{c}_{\ccw}\hat{m}_{1})-\frac{ig_{2}}{\Delta_{2}}(\hat{c}_{\ccw}^{\dagger}\hat{m}_{2}-\hat{c}_{\ccw}\hat{m}_{2}^{\dagger})$, we have $\hat{\mathcal{H}}_{\rm int}+i[\hat{S},\hat{\mathcal{H}}_{0}]=0$. The Baker-Campbell-Hausdorff (BCH) formula gives the following effective Hamiltonian to first order $g_{i}/\Delta_{i}$. Concentrating on the magnon modes, we obtain 
\begin{equation}\label{eq_25}
\begin{split}
\hat{\mathcal{H}}_{\rm eff}=\Omega_1 \hat{m}_{1}^{\dagger}\hat{m}_{1}+\Omega_2\hat{m}_{2}^{\dagger}\hat{m}_{2}+G(\hat{m}_{1}\hat{m}_{2}+\hat{m}_{1}^{\dagger }\hat{m}_{2}^{\dagger}),
\end{split}
\end{equation}
where the effective magnon frequencies are $\Omega_{1}=-g_{1}^{2}/\Delta_{1}+\varpi$ and $\Omega_{2}=-g_{2}^{2}/\Delta_{2}+\omega_{2}$ and $G=g_{1}g_{2}(1/\Delta_{1}+1/\Delta_{2})/2$ denotes the effecive coupling strength between both two magnons. Clearly, the effective Hamiltonian in Eq. (\ref{eq_25}) facilitates the generation of a two-mode squeezed state.

Regarding the Hamiltonian in Eq. (\ref{eq_25}), the coefficient matrix corresponding to the covariance matrix $\bm{\sigma}_{\rm eff}$ can be obtained as follows:
\begin{equation}\label{eq_26}
\begin{split}
\textbf{A}_{2}=\begin{pmatrix}
-\gamma_{1}/2&\Omega_{1}&0&-G\\
-\Omega_{1}&-\gamma_{1}/2&-G&0\\
0&-G&-\gamma_{2}/2&\Omega_{2}\\
-G&0&-\Omega_{2}&-\gamma_{2}/{2}
\end{pmatrix}.
\end{split}
\end{equation}
Under the special conditions of large detuning, the parameters selected in the following discussion continue to satisfy the steady-state conditions according to the Routh-Hurwitz criterion \cite{PhysRevA.35.5288} (see Sec. \ref{Sec.IIIa} for details).

\begin{figure}
\centering
\includegraphics[width=0.43\textwidth]{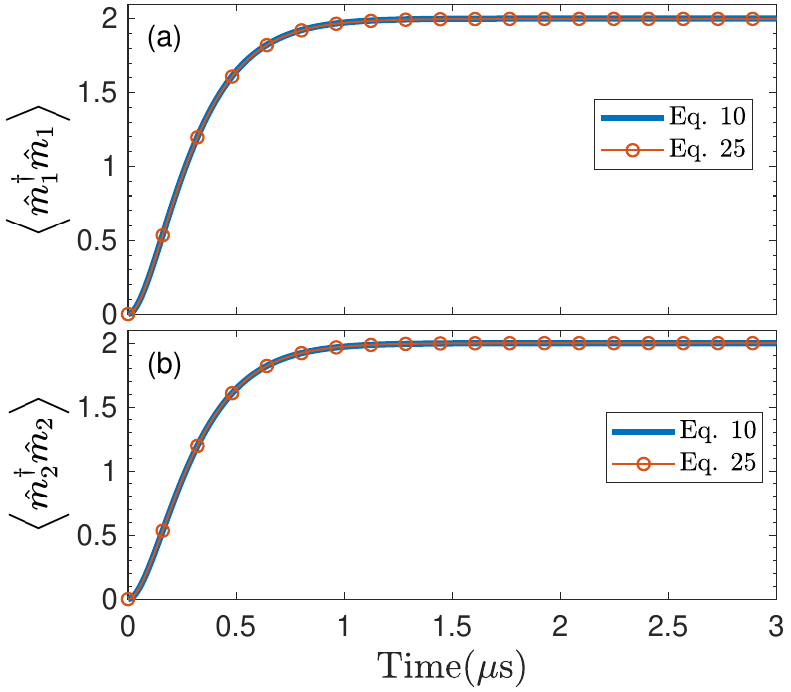}
\caption{The effective Hamiltonian accurately captures the dynamical evolution of the system. The solid blue line and the solid red line with hollow circular markers represent the population of the two magnons, as described by the Hamiltonian Eq. (\ref{eq_10}) and the effective Hamiltonian Eq. (\ref{eq_25}), respectively. The parameters are chosen as follows: $\omega_c/2\pi=10\text{ GHz}$, $\omega_1/2\pi=10.9\text{ GHz}$, $\omega_2/2\pi=9.1\text{ GHz}$, $g_1/2\pi=g_2/2\pi=30\text{ MHz}$, $\omega_{d}/2\pi=20\ \text{GHz}$, $\kappa/2\pi=\gamma_1/2\pi=\gamma_2/2\pi=1\text{ MHz}$. The initial state of the system is that both the cavity and the two magnetons are in a vacuum state.}
\label{fig_06}
\end{figure}

To verify the validity of the approximation in Eq. (\ref{eq_25}), it is essential to compare the system dynamics described by Eqs. (\ref{eq_10}) and (\ref{eq_25}). In the effective Hamiltonian Eq. (\ref{eq_25}), the magnon populations $\hat{m}_{1}$ and $\hat{m}_{2}$ can be determined by the matrix elements of the covariance matrix $\bm{\sigma}_{\rm eff}$ , specifically expressed as $\left\langle\hat{m}_{1}^{\dagger}\hat{m}_{1}\right\rangle =(\sigma_{\rm eff}^{11}+\sigma_{\rm eff}^{22}-1)/2$ and $\left\langle\hat{m}_{2}^{\dagger}\hat{m}_{2}\right\rangle =(\sigma_{\rm eff}^{33}+\sigma_{\rm eff}^{44}-1)/2$, where $\sigma_{\rm eff}^{ii}(i=1,2,3,4)$ denote the diagonal elements of the covariance matrix $\bm{\sigma}_{\rm eff}$ . Similarly, the magnon populations in Hamiltonian Eq. (\ref{eq_10}) can also be represented using the diagonal elements of the covariance matrix. In Fig. \ref{fig_06},
we validate the effectiveness of the Hamiltonian under the required parameter conditions by choosing $\Delta_{1}=\Delta_{2}$ for convenience. The figure clearly shows that, aside from some minor deviations and small oscillations, the population of the two magnon modes corresponding to the two sets of results remains almost identical throughout the entire evolution. Therefore, in the $g_{1}\ll\Delta_{1}$ and $g_{2}\ll\Delta_{2}$ parameter regime, it is reasonable and effective to ignore the higher-order term in Eq. (\ref{eq_10}).

Above, we assumed that the two magnon modes possess identical detuning and coupling strengths. However, in practical YIG sphere-based systems, magnons may exhibit distinct intrinsic frequencies and coupling strengths. Such parameter disparities can significantly influence the effective coupling between the two magnon modes. Here, we systematically analyze the dependence of magnon-magnon entanglement on these parameters to identify the optimal parameter set that maximizes entanglement. To elucidate the effects of unequal magnon frequencies and coupling strengths, we employ the steady-state Lyapunov equation Eq. (\ref{eq_20}) under the effective Hamiltonian Eq. (\ref{eq_25}) to characterize how entanglement evolves with detuning and coupling asymmetry.

\begin{figure}
\centering
\includegraphics[width=0.4\textwidth]{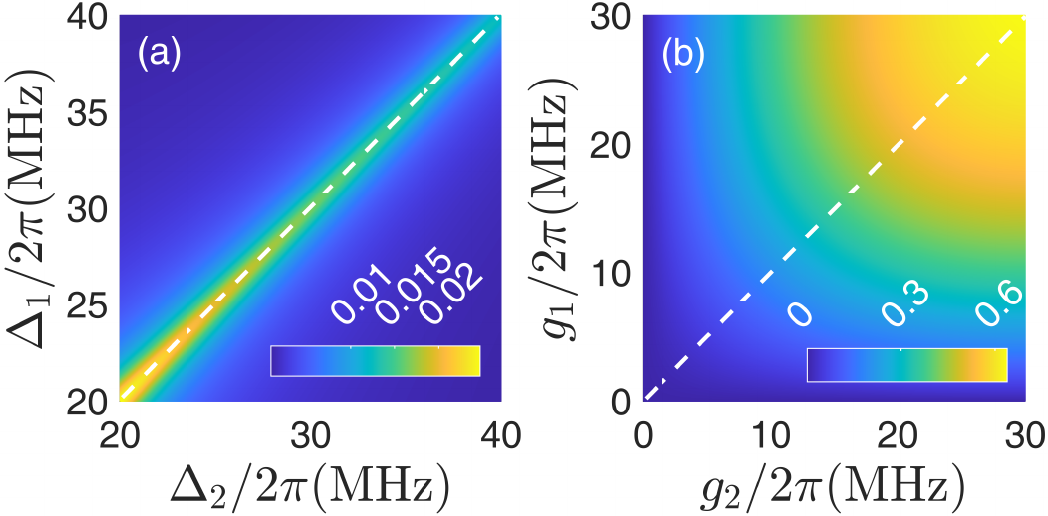}
\caption{The entanglement between the two magnons is examined as follows: (a) The relationship between entanglement and detuning is presented, with the coupling strengths set to $g_{1}/2\pi=g_{2}/2\pi=0.5\text{ MHz}$; (b) The relationship between entanglement and coupling strength is explored with the detuning set to $\Delta_{1}/2\pi=\Delta_{2}/2\pi=0.9\text{ GHz}$. For detailed information on other relevant parameters, please refer to Fig. \ref{fig_03}.}
\label{fig_07}
\end{figure}

In the steady-state entanglement analysis, we first consider the scenario where the coupling strengths of the two modes are equal, i.e., $g_{1}=g_{2}$, with the detunings $\Delta_{1}$ and $\Delta_{2}$ being unconstrained. Based on this, Fig. \ref{fig_07}(a) demonstrates the relationship between entanglement and detuning under constant coupling strength. The results show that as the detuning parameters $\Delta_{1}$ and $\Delta_{2}$ decrease, the entanglement between the two magnon modes increases. It is noteworthy that the entanglement reaches its maximum when  $\Delta_{1}=\Delta_{2}$ and the minimum detuning condition is satisfied under large detuning. Simultaneously, Fig. \ref{fig_07}(b) presents the entanglement between the cavity and two magnon modes under large coupling strengths. Notably, when the detunings of both magnon modes are equalized $\Delta_{1}=\Delta_{2}$, the system exhibits a significantly expanded steady-state region. This extended stabilization window substantially enhances the feasibility of preparing strong magnon-magnon entanglement, as demonstrated in the parameter space of our configuration. Our findings indicate that the entanglement increases with the coupling strength, but once a certain threshold is exceeded, it stabilizes.

In the following, we derive the relationship between entanglement and coupling strengths using analytical solutions. To ensure consistency with the approach used in previous work, we define the covariance matrix as
\begin{equation}\label{eq_27}
\begin{split}
\bm{\sigma}_{\text{eff}}=\begin{pmatrix}
\bm{\Phi}_{\text{eff},\ \hat{m}_{1}}&\bm{\Phi}_{\text{eff},\ \hat{m}_{1}\hat{m}_{2}}\\
\bm{\Phi}_{\text{eff},\ \hat{m}_{1}\hat{m}_{2}}^{\dagger}&\bm{\Phi}_{\text{eff},\ \hat{m}_{2}}
\end{pmatrix}.
\end{split}
\end{equation}
By substituting the elements of the covariance matrix above into Eq. (\ref{eq_18}) and Eq. (\ref{eq_19}), we ultimately obtain the entanglement expression as $E_{N}=-\ln(2\eta)=\ln(1+2G/\sqrt{(\Omega_{1}+\Omega_{2})^2+\gamma^2})$. For clarity, the entanglement measure can be simplified as $E_{N}<\ln(1 + 2G/\sqrt{(\Omega_1 + \Omega_2)^2})$. When $\Delta_{1}=\Delta_{2}$, the right-hand side of the inequality becomes $E_{N}<\ln(1 +2g_1g_2/(g_1^2 +g_2^2))$. It is clear that the maximum entanglement $E_N<\ln2 = 0.6931$ is achieved when $g_1=g_2$. However, as shown in Fig. \ref{fig_07}(b),  maximal entanglement requires not only $g_1=g_2$ but also that $g_1$ and $g_2$ reach their optimal values under the large-detuning condition. This is because larger coupling strengths $g_1$ and $g_2$ further suppress the detrimental effects of dissipation $\gamma$, bring the system closer to the theoretical upper bound of entanglement.

In Fig. \ref{fig_07}, we observe that as the coupling strengths $g_{1}$ and $g_{2}$ increase, the degree of entanglement in the system significantly enhances. Additionally, when the detunings $\Delta_{1}$ and $\Delta_{2}$ decrease, the degree of entanglement in the system also significantly increases. This phenomenon can be attributed to the enhancement of the effective coupling strength $G=g_{1}g_{2}(1/\Delta_{1}+1/\Delta_{2})/2$, which in turn facilitates a greater degree of entanglement between the two magnons. Based on this analysis, we can generate a strong two-mode squeezed state between the two magnons.

From the aforementioned analysis, it is evident that under large detuning, optimal entanglement is achieved when the detuning values are equal. This is because when the detuning is unequal, the parameter range satisfying the stability condition is significantly reduced. Therefore, when the detuning is unequal, and the stability condition is satisfied, the coupling strength decreases significantly, resulting in reduced entanglement. To maximize entanglement, we examine the case of equal detuning. The results indicate that entanglement is maximized when the coupling strengths are equal ($g_1=g_2$) and when the coupling strengths reach their optimal values under the large-detuning condition, with the maximum entanglement value not exceeding $\ln 2$.

\subsection{Experimental feasibility}
\label{Sec.IIId}
To evaluate the experimental feasibility of our two-mode squeezed state generation scheme, we will discuss the experimentally feasible parameters. As mentioned previously, our protocol is implemented in a toroidal cavity. In similar experimental configurations \cite{PhysRevApplied.19.014030,PhysRevApplied.13.044039,PhysRevApplied.16.064066,PhysRevA.105.013711}, magnon modes in toroidal cavities and YIG spheres typically operate in the 6-11 GHz frequency range, with coupling strengths (position-dependent) spanning 0-50 MHz. For this work, we selected frequencies of 8-11 GHz in the toroidal cavity and YIG spheres, with magnon-photon coupling strengths spanning 0-50 MHz. These parameter values align with established experimental capabilities, confirming the scheme's technical viability. The assumed initial state being in the vacuum state can be simply achieved by placing the system at a low bath temperature, e.g., tens of mK.

\section{Conclusion}
\label{Sec.V}

This paper investigates a chiral cavity-magnon system composed of a ring cavity and two YIG spheres, achieving cavity-mediated entanglement and squeezed magnon states. The magnon mode of each YIG sphere
is selectively coupled to one of the two degenerate rotating microwave modes of the toroidal cavity. Building on this framework, we propose two distinct approaches for preparing two-mode squeezed states.

By leveraging a sideband cooling mechanism, the dissipative effects of the cavity field are utilized to cool the Bogoliubov modes of the two magnons to their ground states, thereby realizing squeezed and entangled quantum states. During this process, we observe that the entanglement between the two magnons gradually increases with the coupling ratio $g_{1}/g_{2}$, reaching a maximum before rapidly decreasing. Simultaneously, the entanglement enhances with increasing coupling constant $g_{2}$, and within a specific parameter range, higher cavity dissipation rates $\kappa$ further amplify the entanglement. 

Under large-detuning conditions, the cavity field is adiabatically eliminated to optimize quantum entanglement and squeezing between the magnon modes. We find that the steady-state entanglement between the magnon modes is maximized when the conditions $\Delta_{1}=\Delta_{2}$, $g_{1}=g_{2}$ and the coupling strengths reach their optimal values under the large-detuning condition. The entanglement strength increases with the coupling strengths $g_{1}$ and $g_{2}$, but decreases as the detunings $\Delta_1$ and $\Delta_2$ grow, ultimately producing a maximum entanglement not exceeding $\ln 2$. Through theoretical analysis and numerical simulations, we validate the efficacy of both methods under various experimental parameters, elucidate their impacts on quantum state preparation, and identify optimal squeezing parameters.

The first approach outperforms the second in entanglement generation. Although the second method can enhance quantum entanglement under specific conditions, its maximum entanglement strength is inherently limited by the large-detuning requirement and remains lower than that achievable via the first method. Notably, both approaches circumvent the complexities associated with traditional non-linear mechanisms or external squeezing fields, offering novel pathways for quantum information processing. Future studies could explore the application of these methods in other physical systems and integrate them with emerging quantum technologies to advance the frontiers of quantum information science.

\section*{Acknowledgments}
The authors would like to thank J. Li for valuable discussions. This work is supported by the Innovation Program for Quantum Science and Technology (No. 2023ZD0300700). C.W. was  supported by the National Research Foundation, Singapore and A*STAR under its Quantum Engineering Programme (NRF2021-QEP2-02-P03). G.W. is supported by the Fundamental Research Funds for the Central Universities (Grant No. 2412020FZ026) and the Key Laboratory of Advanced Optical Manufacturing Technologies of Jiangsu Province, Soochow University (Grant No. KJS2336).
\appendix
\renewcommand{\appendixname}{APPENDIX}

\section{Stable condition}
\label{app_A}

The necessary conditions for stability can be determined using the Routh-Hurwitz criterion \cite{PhysRevA.35.5288}, namely, the system becomes stable and attains its steady state only when the real parts of $\textbf{A}$ are negative. To achieve this, we need to derive the characteristic equation of matrix A, specifically $|\lambda I-\textbf{A}|=0$, which will yield the desired characteristic polynomial 
\begin{equation}\label{eq_A1}
a_{0}\lambda^{6}+a_{1}\lambda^{5}+a_{2}\lambda^{4}+a_{3}\lambda^{3}+a_{4}\lambda^{2}+a_{5}\lambda+a_{6}=0,
\end{equation}
where
\begin{widetext}
\begin{equation}\label{eq_A2}
\begin{split}
a_{0}=&1,\\
a_{1}=&\gamma_{1}+\gamma_{2}+\kappa,\\
a_{2}=&\frac{1}{4}(\gamma_{1}^{2}+\gamma_{2}^2+\kappa^{2})+\gamma_{1}\gamma_{2}+\gamma_{1} \kappa+\gamma_{2}\kappa-2g_{1}^2+2g_{2}^{2}+\omega_{2}^{2}+\omega_{c}^{2}+\varpi^{2},\\
a_{3}=&\frac{1}{4}(\gamma_{1}^{2}\gamma_{2}+\gamma_{1}^{2}\kappa +\gamma_{1}\gamma_{2}^{2}+4\gamma_{1}\gamma_{2}\kappa+4\omega_{c}^2(\gamma_{1}+\gamma_{2})+\gamma_{1}\kappa^{2}+4\gamma_{1}\omega_{2}^2+\gamma_{2}^{2}\kappa+\gamma_{2}\kappa^{2}+4\gamma_{2}\varpi^{2}-4g_{1}^2 (\gamma_{1}+2\gamma_{2}+\kappa)\\
&+4g_{2}^{2}(2\gamma_{1}+\gamma_{2}+\kappa)+4\kappa\omega_{2}^{2}+4\kappa\varpi^{2}),\\
a_{4}=&\frac{1}{16}(16g_{1}^{4}+16g_{2}^{4}+\gamma_{1}^{2}\gamma_{2}^{2}+4\gamma_{1}^{2}\gamma_{2}\kappa+4\gamma_{1}\gamma_{2}^{2}\kappa+\gamma_{1}^{2}\kappa^{2}+4\gamma_{1}\gamma_{2}\kappa^{2}+\gamma_{2}^{2}\kappa^{2}+4\gamma_{2}^{2}\varpi^{2}+16\gamma_{2}\kappa\varpi^{2}+4\kappa^{2}\varpi^{2}+4\gamma_{1}^{2}\omega_{2}^{2}\\
&+16\gamma_{1}\kappa\omega_{2}^{2}+4\kappa^{2}\omega_{2}^{2}+16\omega_{2}^{2}\varpi^{2}+4\omega_{c}^{2}(\gamma_{1}^{2}+4\gamma_{1}\gamma_{2}+\gamma_{2}^{2}+4(\omega_{2}^{2}+\varpi^{2}))-8g_{1}^{2}(4g_{2}^{2}+2\gamma_{1}\gamma_{2}+\gamma_{2}^2+\gamma_{1}\kappa+2\gamma_{2}\kappa\\
&+4\omega_{2}^{2}+4\omega_{c}\varpi)+8g_{2}^{2}(\gamma_{1}^{2}+\gamma_{2}\kappa+ 2\gamma_{1}(\gamma_{2}+\kappa)-4\omega_{2}\omega_{c}+4\varpi^{2})),\\
a_{5}=&\frac{1}{16}(16g_{2}^{4}\gamma_{1}+16g_{1}^{4}\gamma_{2}+\gamma_{2}\kappa(\gamma_{1}\gamma_{2}\kappa+\gamma_{1}^{2}(\gamma_{2}+\kappa)+4(\gamma_{2}+\kappa)\varpi^{2})+4\omega_{2}^{2}\kappa(\gamma_{1}(\gamma_{1}(\gamma_{1}+\kappa)+4\varpi^{2})+4\omega_{c}^{2}(\gamma_{1}\gamma_{2}(\gamma_{1}\\
&+\gamma_{2})+4\gamma_{2}\varpi^{2}+4\gamma_{1}\omega_{2}^{2})-4g_{1}^{2}(4g_{2}^{2}(\gamma_{1}+\gamma_{2})+\gamma_{2}^{2}\kappa+4\kappa
\omega_{2}^{2}+\gamma_{1}(\gamma_{2}^{2}+2\gamma_{2}\kappa+4\omega_{2}^{2})+8\gamma_{2}\varpi\omega_{c})+4g_{2}^{2}(\gamma_{1}^{2}(\gamma_{2}\\
&+\kappa)+4\varpi^{2}(\gamma_{2}+\kappa)+2\gamma_{1}(\gamma_{2}\kappa-4\omega_{2}\omega_{c}))),\\
a_{6}=&\frac{1}{64}(-8g_{2}^{2}(4g_{1}^{2} (\gamma_{1}\gamma_{2}-4\omega_{2}\varpi)-(\gamma_{1}^{2}+4\varpi^{2})(\gamma_{2}\kappa-4\omega_{2}\omega_{c}))+(\gamma_{2}^{2}+4\omega_{2}^{2})((\gamma_{1}^{2}+4\varpi^{2})(\kappa^{2}+4\omega_{c}^{2})-8g_{1}^{2}(\gamma_{1}\kappa\\
&+4\omega_{c}\varpi ))+16g_{2}^4(\gamma_{1}^{2}+4\varpi^{2}))+16g_{1}^{4}.
\end{split}
\end{equation}
\end{widetext}

Based on the coefficients $a_{k}$ ($k = 1,2,3,4,5,6$), six Hurwitz matrices can be constructed, where the $k$-th matrix has a dimension of $k \times k$ ($1 \leq k \leq 6$). The elements of these matrices are determined by the following condition \cite{PhysRevA.109.022601}:
\begin{equation}\label{eq_A3}
\begin{split}
\mathfrak{T}_{i j}^{k}=\left\{\begin{matrix}
 0, & 2 i-j<0   & \text { or } \quad 2 i-j>k\\
 a_{2 i-j}, & \text {otherwise } &
\end{matrix}\right.
\end{split}
\end{equation}
where $1 \le i, j \le k$. The stability condition of the system requires that all determinants of the Hurwitz matrices be positive, i.e.,
\begin{equation}\label{eq_A4}
\begin{split}
\forall \text{det}[\mathfrak{T}^{k}]>0.
\end{split}
\end{equation}

Based on Eq. (\ref{eq_A4}), the stability of the system is guaranteed if the following conditions are satisfied:
\begin{equation}\label{eq_A5}
\begin{split}
& a_{k}>0, \quad k=1,\cdots,6;\\
& a_{1}a_{2}>a_{3},\quad 
a_{1}a_{2}a_{3}>a_{3}^{2}+a_{1}^{2}a_{4},\\ &T_{1}>T_{2}, \quad T_{3}>T_{4},
\end{split}
\end{equation}
 where
\begin{equation}\label{eq_A6}
\begin{split}
T_{1}= &(a_{1}a_{4}-a_{5})(a_{1} a_{2}a_{3}-a_{3}^{2}-a_{1}^{2} a_{4}), \\
T_{2}= & a_{1} a_{5}^{2}+a_{5}(a_{1}a_{2}-a_{3})^{2} \\
T_{3}= & a_{1}^{2}a_{6}(2 a_{2} a_{5}+a_{3} a_{4})+a_{3}^{3} a_{6} +a_{1}a_{2}a_{3}a_{4}a_{5}\\
&+a_{5}^{2}(2a_{1} a_{4}+a_{2}a_{3}), \\
T_{4}= &a_{1}^{2}(a_{1} a_{6}^{2}+a_{4}^{2} a_{5})+a_{5}^{3}+a_{4}a_{5} a_{3}^{2}\\
&+a_{1}(a_{2}a_{6} a_{3}^{2}+3a_{3}a_{5} a_{6}+a_{2}^{2}a_{5}^{2}) .
\end{split}
\end{equation}
Eq. (\ref{eq_A5}) ensures the stability of the system. Due to the complexity of these analytical results, we employ numerical methods in the main text to identify parameters that simultaneously satisfy experimental constraints and guarantee system stability.

\bibliography{ref}
\vspace{8pt}
\end{document}